\documentclass[prd,12pt,nofootinbib]{revtex4-1}
\usepackage{amsmath}
\usepackage{commath}
\usepackage{graphicx}
\usepackage{amssymb}
\usepackage{subfigure}
\usepackage{bm}
\usepackage{slashed}
\usepackage{tikz}
\usepackage{braket}
\usepackage[labelsep=space]{caption}
\usepackage[capitalise]{cleveref}

\def\process{\mbox{ $B \to K^{*} \nu \bar \nu$ }}

\def\be{\begin{equation}}
\def\ee{\end{equation}}	
\def\bea{\begin{eqnarray}}
\def\eea{\end{eqnarray}}
\def\d{\mathrm{d}}
\def\beq{\begin{equation}}
\def\eeq{\end{equation}}

\def\nnb{\nonumber}

\def\rar{\rightarrow}
\def\nnb{\nonumber}

\def\ba{\begin{array}}
	\def\ea{\end{array}}

\def\tepm{$B \rar K^* \mu^+ \mu^-$}

\begin{document}
\title{Probing transition form factors in the rare $B\to K^*\nu\bar\nu$ decay}

\author{Mohammad Ahmady}
\email{mahmady@mta.ca}
\affiliation{\small Department of Physics, Mount Allison University, \mbox{Sackville, New Brunswick, Canada, E4L 1E6}}
\author{Alexandre Leger}
\email{azleger@mta.ca}
\affiliation{\small Department of Physics, Mount Allison University, \mbox{Sackville, New Brunswick, Canada, E4L 1E6}}
\author{Zoe McIntyre}
\email{zxmcintyre@mta.ca}
\affiliation{\small Department of Physics, Mount Allison University, \mbox{Sackville, New Brunswick, Canada, E4L 1E6}}
\author{Alexander Morrison}
\email{ahmorrison@mta.ca}
\affiliation{\small Department of Physics, Mount Allison University, \mbox{Sackville, New Brunswick, Canada, E4L 1E6}}
%\author{Ryan MacGibbon}
%\email{ryan_macgibbon@me.com}
%\affiliation{\small Department of Physics, Acadia University,
%	\mbox{Wolfville, Nova-Scotia, Canada, B4P 2R6}}
\author{Ruben Sandapen}
\email{ruben.sandapen@acadiau.ca}
\affiliation{\small Department of Physics, Acadia University,
	 \mbox{Wolfville, Nova-Scotia, Canada, B4P 2R6}}
\affiliation{\small Department of Physics, Mount Allison University, \mbox{Sackville, New Brunswick, Canada, E4L 1E6}}
%\affiliation{D\'epartement de Physique et d'Astronomie, Universit\'{e} de Moncton, \mbox{Moncton, New Brunswick, Canada, E1A 3E9}}

\begin{abstract}
 We compare the Standard Model (SM) predictions for the differential branching ratio of the rare $B\to K^*\nu\bar\nu$ decays using $B \to K^*$ form factors obtained from holographic light-front QCD (LFHQCD) and Sum Rules (SR) Distribution Amplitudes.  For the total branching ratio, we predict $\mathcal{BR}(\process)_{\rm LFHQCD}=(6.36^{+0.59}_{-0.74})\times 10^{-6}$ and $\mathcal{BR}(\process)_{\rm SR}=(8.14^{+0.16}_{- 0.17})\times 10^{-6}$.  More interestingly, we find that the two model predictions for the differential branching ratio are sufficiently different at low momentum transfer, so that future measurements at Belle II may be able to discriminate between them. We also confirm numerically that the $K^*$ longitudinal polarization fraction $F_L$ has little sensitivity to the non-perturbative form factors and is thus an excellent observable to probe New Physics signals.  We predict $F_L=0.40^{+0.02}_{-0.01}(0.41\pm 0.01)$ using LFHQCD (SR).

\end{abstract}


\maketitle
%%%%%%%%%%%%%%%%%%%%%%%%%%%%%%%%%%%%%%%%%%
\section{Introduction}
 The flavor changing neutral current (FCNC) $b\to s$ transition has been at the focus of extensive experimental and theoretical investigations.  This is due to the fact that, among other things, this rare transition is sensitive to new physics (NP) beyond the Standard Model (SM).  Many observables for the dileptonic \tepm decay have already been measured and the precision of the experimental data is expected to improve significantly in the near future. On the other hand, the rare decay \process has not yet been measured experimentally and it is challenging to do so, as both leptons are detector eluding neutrinos. Only the upper bounds on the branching ratio ($\mathcal{BR}$) are known and the most ones are set by the Belle Colaboration \cite{Lutz:2013ftz}:
 \bea
 \mathcal{BR}(B^+\to K^{*+}\nu\bar{\nu})&<& 4.0\times 10^{-5}\;\; (90\% \;\; {\rm CL})\;\; ,\nnb\\
 \mathcal{BR}(B^0\to K^{*0}\nu\bar{\nu})&<& 5.5\times 10^{-5}\;\; (90\% \;\; {\rm CL})\;\; .
 \label{upperbound}
 \eea
 With the advent of Super-B facilities, the prospects of measuring these branching ratios in the near future are good.   The Belle-II experiment, with an integrated luminosity $50\; {\rm ab}^{-1}$ that
 is expected to be collected by 2023, a measurement of the SM $\mathcal{BR}$s
 with $30\%$ precision is expected \cite{Aushev:2010bq}. Therefore, it is appropriate to have a closer look at this decay in order to motivate and  further guide experimental
 efforts to measure the $\mathcal{BR}$s and related observables. Theoretically, the presence of only one operator in the effective Hamiltonian for the $b\to s\nu\bar{\nu}$ transition makes \process much less susceptible to hadronic uncertainty due to sensitivity to a minimal number of form factors. Moreover, this decay process does not suffer from additional uncertainties beyond the form factors, such as those that plague the $b \to s\ell^+{\ell}^-$ transitions due to the breaking of factorization caused by photon exchange.  Indeed, for the \process transition, factorization holds exactly, so a measurement of the decay rate would allow in principle to measure the form factors. This distinction also leads to the fact that, in contrast to \tepm decays,
 the isospin asymmetries of the decays with neutrinos in the final state vanish identically, so the
 branching ratio of the $B^0$ and $B^\pm$ decays only differ due to the lifetime differences, i.e. $\mathcal{BR}(B^+\to K^{*+}\nu\bar{\nu})/\mathcal{BR}(B^0\to K^{*0}\nu\bar{\nu})=\tau_{B^+}/\tau_{B^0}$ is valid in the SM and beyond.

In this paper, we calculate the differential $\mathcal{BR}$ as well as the $K^*$ longitudinal polarization fraction for \process decay.  The form factors parameterizing the $B\to K^*$ hadronic matrix elements are derived via light-cone sum rules (LCSR)\footnote{LCSR form factors are accurate for low-to-intermediate $q^2$.} in which the required Distribution Amplitudes (DAs) for $K^*$ are obtained from the holographic light-front wavefunctions (LFWFs) for vector mesons\cite{Ahmady:2014sva} and from QCD Sum Rules \cite{Straub:2015ica}.  Successful predictions of diffractive $\rho$  and $\phi$ meson electro-production at HERA \cite{Adloff:1999kg,Aid:1996bs,Breitweg:1997ed,Chekanov:2005cqa,Collaboration:2009xp,Chekanov:2007zr} using holographic LFWFs, motivates us to use these alternative DAs in our calculation of form factors\cite{Forshaw:2012im,Ahmady:2016ujw}.  The $B\rar (\rho ,\; K^*)$ transition form factors using holographic $\rho$ and $K^*$ DAs have previously been used to calculate the differential decay rate of semileptonic $B\rar\rho\ell\nu$\cite{Ahmady:2013cga} and dileptonic \tepm \cite{Ahmady:2014sva} as well as the isospin asymmetry \cite{Ahmady:2014cpa} and resonance effects\cite{Ahmady:2015fha} in the latter decay.  In \cite{Ahmady:2015fha}, we also found that our predictions for $\mathcal{BR}$(\tepm ) are not very different at low-to-intermediate $q^2$ when using SR or holographic DAs (see Fig 5 of \cite{Ahmady:2015fha}.).  We shall see that the situation is different for $\mathcal{BR}$(\process ).

%%%%%%%%%%%%%%%%%%%%%%%%%%%%%%%%%%%%%%%%%%%%%%%%%%%%%%%%%%%
\section{The effective Hamiltonian}

\begin{figure}
	\centering
	\subfigure[\mbox{ }Penguin diagram]{\includegraphics[width=0.3\textwidth]{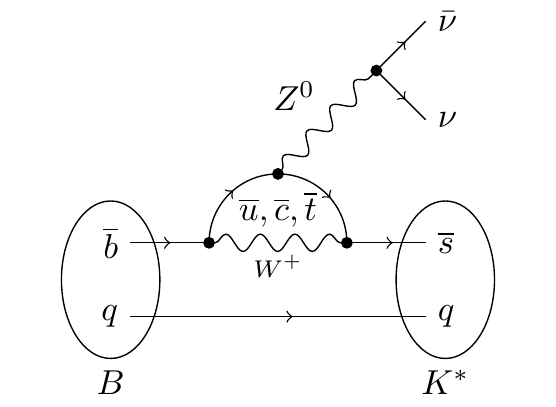}}
	\subfigure[\mbox{ }Box diagram]{\includegraphics[width=0.3\textwidth]{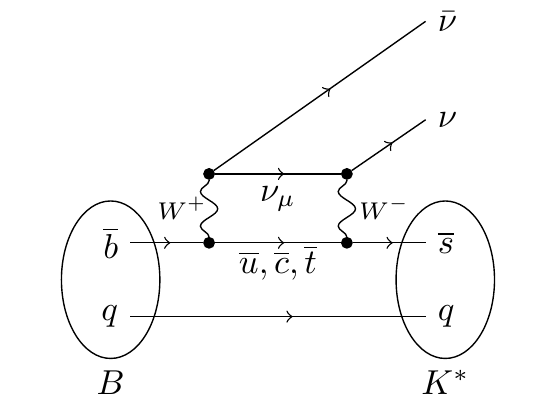}}
	\caption{Feynman diagrams of the principal contributions to the $B\to K^{*}\nu\bar\nu$ decay.}
	\label{fig:feynmandiagrams}
\end{figure}

%%%%%%%%%%%%%%%%%%%%%%%%%%%%%%%%%%%%%%%%%%%%%%%%%%%%%%%%%%%
%%%%%%%%%%%%%%%%%%%%%%%
The effective Hamiltonian for FCNC transition $b\to s\nu\bar \nu$ in the SM is given as\cite{Buras:2014fpa}
\be
\mathcal{H}=-\frac{4G_F}{\sqrt{2}}V_{tb}V_{ts}^{*}C_{L}\mathcal{O}_L + h.c. \; ,
\label{effectivehamiltonian}
\ee
where the only contributing operator $\mathcal{O}_{L}$ is defined as:
\be
\mathcal{O}_{L}=\frac{e^2}{8\pi^2}(\bar{s}\gamma_{\mu}P_Lb)(\bar{\nu}\gamma^{\mu}P_L\nu)\; , 
\ee
where $P_L=(1-\gamma_5)/2$ is the left-handed projection operator and $C_{L}$ is the Wilson coefficient given by
\be
C_{L}=\frac{-X(x_t) }{s^2_w} \; ,
\ee
with $s_w^2=\sin^2 \theta_w\sim 0.23$ ($\theta_w$ is the weak angle) and $x_t=m_t^2/m_W^2$.   The leading-order (LO) contributions to $X(x_t)$ can be written as:
\be
X_0(x_t)=C_0(x_t)-4B_0(x_t) \;\; ,
\ee
where the effective vertex functions $B_0$ and $C_0$ represent the box and penguin diagrams shown in Figure (\ref{fig:feynmandiagrams}).  Explicitly,
\bea
B_0(x)&=& \frac{x}{4(x-1)^2}\ln (x)-\frac{x}{4(x-1)}\; ,\\
C_0(x) &=& \frac{3x^2+2x}{8(x-1)^2}\ln (x)+\frac{x^2-6x}{8(x-1)}\; .
\eea

Using $m_t=173.21\pm 0.51\pm 0.71$ GeV and $m_W=80.385\pm 0.015$ GeV \cite{Agashe:2014kda}, one obtains $X_0=1.59\pm 0.01$ to LO.  Including the next-to-leading order (NLO) QCD corrections along with the two-loop electroweak contributions reduces $X_0$ by about $10\%$ to $X =1.469\pm0.017$ \cite{Misiak:1999yg,Buchalla:1998ba,Brod:2010hi}.

The transition $B\to K^*$ via the effective Hamiltonian given by Eq. (\ref{effectivehamiltonian}) is parametrized by four form factors:  
\bea
\langle K^* (k,\varepsilon)|\bar{s} \gamma^\mu(1-\gamma^5 )b | B(p) \rangle &=& \frac{2i V(q^2)}{m_B + m_{K^*}} \epsilon^{\mu \nu \rho \sigma} \varepsilon^*_{\nu} k_{\rho} p_{\sigma} -2m_{K^*} A_0(q^2) \frac{\varepsilon^* \cdot q}{q^2} q^{\mu}  \nonumber \\
&-& (m_B + m_{K^*}) A_1(q^2) \left(\varepsilon^{\mu *}- \frac{\varepsilon^* \cdot q q^{\mu}}{q^2} \right) \nonumber \\
&+& A_2(q^2) \frac{\varepsilon^* \cdot q}{m_B + m_{K^*}}  \left[ (p+k)^{\mu} - \frac{m_B^2 - m_{K^*}}{q^2} q^{\mu} \right] \; ,
\label{kstarformfactors}
\eea
where $\varepsilon$ is the polarization $4$-vector of the $K^*$.  The form factors $A_0,\; A_1,\; A_2$ and $V$ are sensitive to nonperturbative QCD and therefore one should resort to a specific model to calculate them.  In a previous paper, we calculated $B\rar K^*$ transition form factors by using the light-cone sum rules (LCSR) along with holographic distribution amplitudes (DAs) obtained from holographic light-front QCD\cite{Ahmady:2014sva}.  In this paper, we also estimate the uncertainty in the form factors resulting from the uncertainties due to quark masses.

%%%%%%%%%%%%%%%%%%%%%%%%%%%%%%%%%%%%%%%%%
\section{Distribution Amplitudes for the $K^*$}
We now proceed to predict the twist-$2$ DAs $\phi_{K^*}^{\parallel ,\perp} (x,\; \mu )$ by writing them in terms of the light-front wavefunctions for $K^*$ \cite{Ahmady:2013cva}:
\begin{equation}
	f_{K^*} \phi_{K^*}^\parallel(x,\mu) =\sqrt{\frac{N_c}{\pi}} \int \d
	b \mu
	J_1(\mu b) \left[1 + \frac{m_{\bar{q}} m_{s} -\nabla_b^2}{M_{K^*}^2 x(1-x)}\right] \Psi_L(x, \zeta) \;,
	\label{phiparallel-phiL}
\end{equation}
and 
\begin{equation}
	f_{K^*}^{\perp}(\mu) \phi_{K^*}^\perp(x,\mu) =\sqrt{\frac{N_c }{2 \pi}} \int \d
	b \mu
	J_1(\mu b) [m_s - x(m_s-m_{\bar{q}})] \frac{\Psi_T(x,\zeta)}{x(1-x)} \; ,
	\label{phiperp-phiT}
\end{equation}
where $f_{K^*}$ and $f_{K^*}^\perp$ are the longitudinal and transverse coupling constants, respectively. $\mu$ is the nonperturbative hadronic scale and the above expressions are valid at $\mu \sim 1$ GeV, which is a scale representing transition from perturbative to non-perturbative regimes.  $\Psi_{\lambda}(x,\zeta)$ are holographic meson wavefunctions obtained by solving the holographic light-front Schr\"odinger Equation for mesons \cite{Brodsky:2014yha}. Explicitly,

\be  
\Psi_{\lambda} (x,\zeta) = \mathcal{N}_{\lambda} \sqrt{x (1-x)}  \exp{ \left[ -{ \kappa^2 \zeta^2  \over 2} \right] }
\exp{ \left[ -{(1-x)m_s^2+ xm_{\bar{q}}^2 \over 2 \kappa^2 x(1-x) } \right]} \;,
\label{hwf}
\ee
where $\lambda =L,\; T$ denotes the polarization and $\zeta=\sqrt{x\bar x}b$ is the so-called holographic variable \cite{Brodsky:2014yha}.  The polarization-dependent normalization constant ${\mathcal N}_{\lambda}$ by requiring that \cite{Forshaw:2012im}
\be
\sum_{h,\bar{h}} \int {\mathrm d}^2 b \, {\mathrm d} x |
\Psi^{K^*, \lambda} _{h, {\bar h}}(x, b)|^{2} = 1 \;,
\label{normalisation}
\ee
where the helicity-dependent wavefunctions in Eq. \ref{normalisation} are given by \cite{Ahmady:2013cva,Ahmady:2016ujw}

\be
\Psi^{K^*,L}_{h,\bar{h}}(x,b) =  \frac{1}{2} \delta_{h,-\bar{h}}  \bigg[ 1 + 
{m_{\bar{q}} m_{s} -  \nabla_b^{2}  \over x(1-x)M^2_{K^*} } \bigg] \Psi_L(x, \zeta) \;  ,
\label{mesonL}
\ee
and
\be \Psi^{K^*, T}_{h,\bar{h}}(x,b) = \pm \bigg[  i e^{\pm i\theta_{b}}  ( x \delta_{h\pm,\bar{h}\mp} - (1-x)  \delta_{h\mp,\bar{h}\pm})  \partial_{b}+ [xm_{\bar{q}} + (1-x) m_s]\delta_{h\pm,\bar{h}\pm} \bigg] {\Psi_T(x, \zeta) \over 2 x (1-x)} \; .
\label{mesonT}
\ee

 In Eq.\eqref{hwf}, $\kappa$ is the fundamental confinement scale \cite{Brodsky:2014yha} that emerges in light-front holography. Spectroscopic data indicate that $\kappa=0.55$ GeV for light vector mesons and a similar value, $\kappa=0.54$ GeV, is also favoured by the data for diffractive electroproduction of $\rho$ and $\phi$ vector mesons \cite{Forshaw:2012im,Ahmady:2016ujw}. We shall therefore use the latter value here. As for the quark masses $m_{\bar{q}/s}$, we shall fix them here in order to fit the experimentally measured decay constant $f_{K^*}$ and we also check that our prediction for the ratio of transverse to longitudinal coupling, $f_{K^*}^{\perp}/f_{K^*}$ is in reasonable agreement with lattice predictions. 
 
The longitudinal and transverse couplings are given by \cite{Ahmady:2013cva,Ahmady:2016ujw}
\bea
f_{K^*} &=&  {\sqrt \frac{N_c}{\pi} }  \int_0^1 {\mathrm d} x  \left[ 1 + { m_{\bar{q}}m_s-\nabla_{b}^{2} \over x (1-x) M^{2}_{K^*} } \right] \left. \Psi_L(\zeta, x) \right|_{\zeta=0}\; ,
\label{fvL}
\eea
and  
\begin{equation}
f_{K^*}^{\perp}(\mu) =\sqrt{\frac{N_c}{2\pi}}  \int_0^1 {\mathrm d} x (xm_{\bar{q}} +(1-x)m_s)  \int {\mathrm d} b  \; \mu J_1(\mu b)  \frac{\Psi_T(\zeta,x)}{x(1-x)} \; ,
\label{fvT}
\end{equation}
respectively, which are obtained from the normalization condition on DAs, i.e.
\be
\int \phi_{K^*}^{\parallel ,\perp}(x,\; \mu ) dx=1\; .
\ee
 Our predictions are shown in Table \ref{tab:decay}. As can be seen, different sets of quark masses can be used to fit the measured decay constant with the larger quark masses being preferred in order to approach the lattice data for the ratio $f_{K^*}^{\perp}(\mu)/f_{K^*}$. Guided by our predictions in Table \ref{tab:decay}, we shall use $m_{\bar{q}}=(195 \pm 55)$ MeV and $m_s=(300 \pm 20)$ MeV in this paper. 

\begin{table}[h]
%\begin{center}
%\textbf{AdS/QCD predictions for the decay constants of $K^*$}
\[
\begin{array}
[c]{|c|c|c|c|c|c|c|}\hline
\mbox{Approach}&\mbox{Scale}~ \mu  &m_{\bar{q}} \mbox{[MeV]} & m_s \mbox{[MeV]} &f_{K^*} \mbox{[MeV]} &f_ {K^*}^{\perp} (\mu) \mbox{[MeV]}&f_{K^*}^{\perp}/f_{K^*} (\mu)\\ \hline
%\rho & 0.548 & 0.140&0.140&0.214&0.135&0.63 \\ \hline
 \mbox{LFHQCD} & \sim 1~\mbox{GeV} & 140 & 280 & 200  & 118 & 0.59 \\ \hline
\mbox{LFHQCD} & \sim 1~\mbox{GeV} & 195 & 300 & 200  & 132 & 0.66 \\ \hline
 \mbox{LFHQCD} & \sim 1~\mbox{GeV} & 250 & 320 & 200  & 142 & 0.71 \\ \hline \mbox{Experiment}  & &  & &205\pm 6\footnote{From $\Gamma(\tau^- \to K^{*-} \nu_{\tau})$}  & & \\ \hline
% \mbox{SR}  & 1 ~\mbox{GeV}& 220 \pm 5&  185 \pm 10& 0.82 \pm 0.06\\ \hline
%\mbox{SR}  & 2 ~\mbox{GeV}& 220 \pm 5& 162 \pm 9 & 0.73 \pm 0.04\\ \hline
\mbox{Lattice}  & 2 ~\mbox{GeV}& & & & &0.780 \pm 0.008 \\ \hline
\mbox{Lattice}  & 2 ~\mbox{GeV}& &  & &  & 0.74 \pm 0.02\\ \hline
\end{array}
\]
%\end{center}
\caption {Comparison between LFHQCD predictions for the decay constant of the $K^*$ meson with experiment \cite{Beringer:1900zz}, and the ratio of couplings with lattice \cite{Becirevic:2003pn,Braun:2003jg} data.}
\label{tab:decay}
\end{table}

We can now compare the holographic DAs with those obtained using QCD Sum Rules. Note that Sum Rules predict the moments of the DAs: 
\begin{equation}
\langle \xi_{\parallel, \perp}^n \rangle_\mu = \int \d x \; \xi^n \phi_{K^*}^{\parallel,\perp} (x, \mu)  
\end{equation}
and that only the first two moments are available in the standard SR approach \cite{Ball:2007zt}. The twist-$2$ DA are then reconstructed as a Gegenbauer expansion
\begin{equation}
\phi_{K^*}^{\parallel,\perp}(x, \mu) = 6 x \bar x \left\{ 1 + \sum_{j=1}^{2}
a_j^{\parallel,\perp} (\mu) C_j^{3/2}(2x-1)\right\} \;. 
\label{phiperp-SR}
\end{equation}
where $C_j^{3/2}$ are the Gegenbauer polynomials and the coeffecients $a_j^{\parallel,\perp}(\mu)$ are related to the moments $\langle \xi_{\parallel,\perp}^n \rangle_\mu$ \cite{Choi:2007yu}. These moments and coefficients are determined at a starting scale $\mu=1$ GeV and can then  be evolved perturbatively to higher scales \cite{Ball:2007zt}.

%%%%%%%%%%%%%%%%%%%%%%%%%%%%%%%%
\begin{figure}[htbp]
	\begin{subfigure}{}
	\centering
	\includegraphics[width=0.3\textwidth]{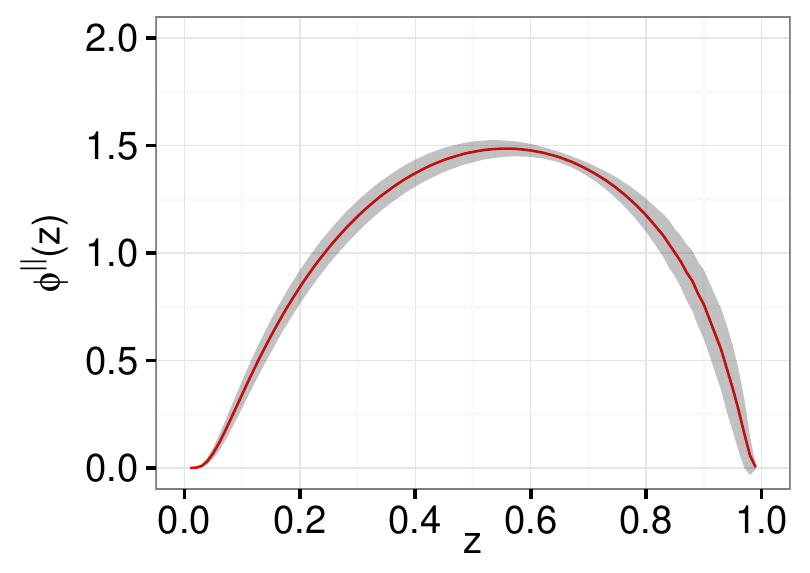}
	%\caption{1a}
	\label{adsdapara}
\end{subfigure}
\hspace{.1cm}
\begin{subfigure}{}
	\centering
	\includegraphics[width=0.3\textwidth]{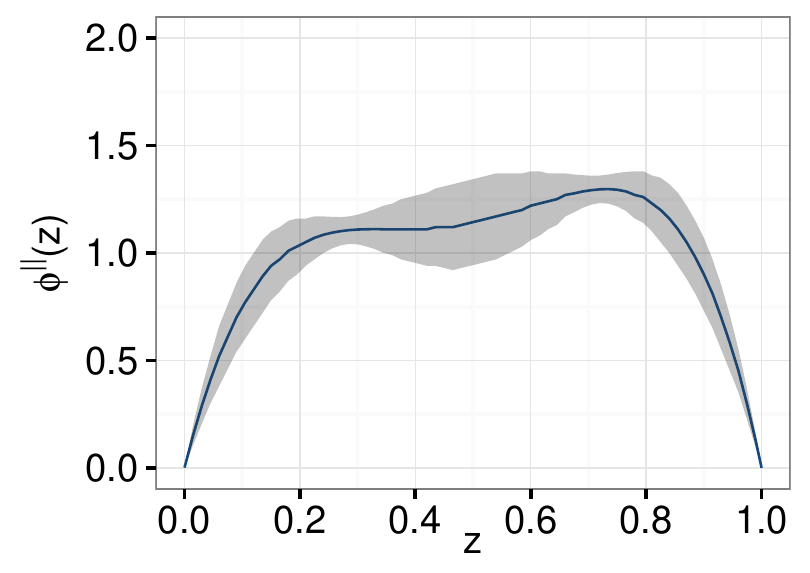}
	%\caption{1b}
	\label{srdapara}
\end{subfigure}\\
\begin{subfigure}{}
	\centering
	\includegraphics[width=0.3\textwidth]{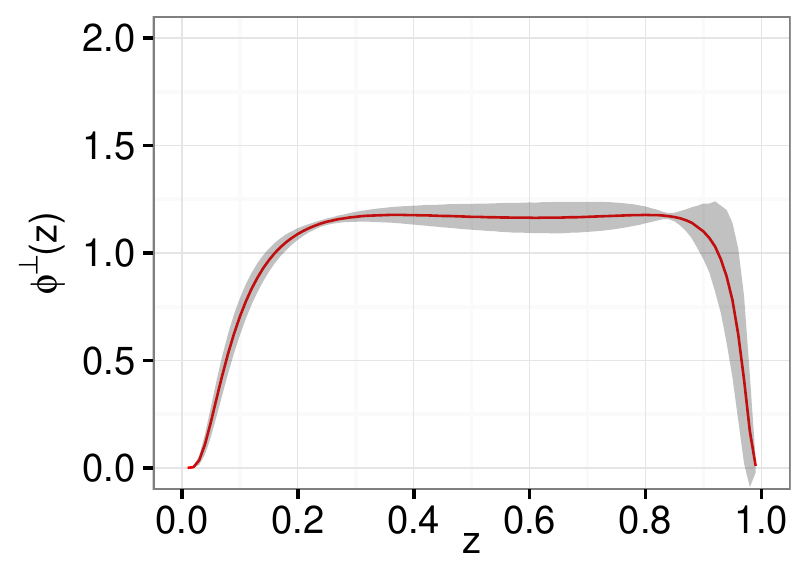}
	%\caption{1c}
	\label{adsdaperp}
	\end{subfigure}
	\begin{subfigure}{}
	\centering
	\includegraphics[width=0.3\textwidth]{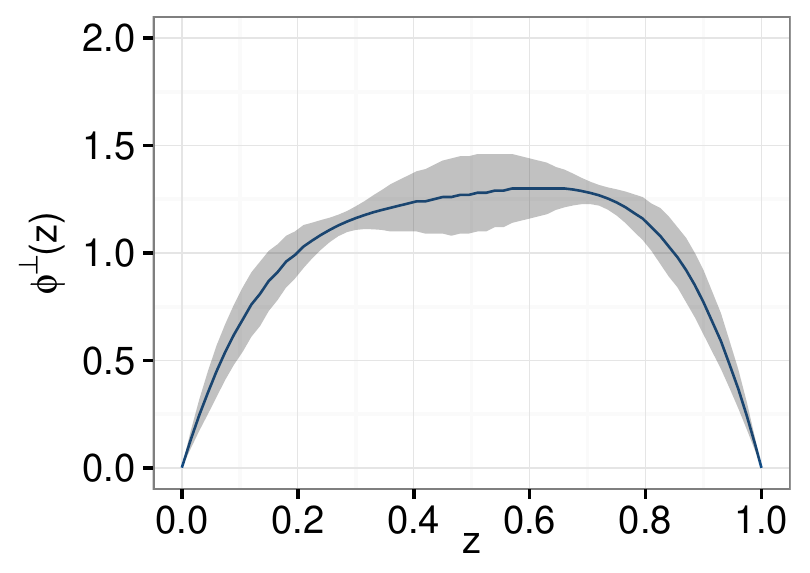}
	%\caption{1c}
	\label{srperp}
	\end{subfigure}
	\caption{Twist-2 DAs predicted by LFHQCD (graphs on the left) and SR (graphs on the right).  The uncertainty band is due to the variation of the quark masses for AdS/QCD and the error bar on Gegenbauer coefficients for SR.}
	\label{das}
\end{figure}
Figure \ref{das} shows twist-2 DAs $\phi^{\parallel ,\perp}(z,\; \mu=1\; {\rm GeV})$ for the $K^*$ vector meson obtained from Eqs. \ref{phiparallel-phiL} and \ref{phiperp-phiT} as compared to SR predictions as given by Eq. \ref{phiperp-SR}.  The uncertainty band for holographic DAs are due to different sets of ($m_s,\; m_{\bar q}$) that fit the measured decay constant as shown in Table \ref{tab:decay}.  The error band in SR DAs are the result of the uncertainties in the Gegenbauer coefficients, ie. $a_1^\parallel =0.06\pm 0.04$, $a_2^\parallel =0.16\pm 0.09$ for $\phi^\parallel(z,\; \mu=1\; {\rm GeV})$ and $a_1^\perp =0.04\pm 0.03$, $a_2^\perp =0.10\pm 0.08$ for $\phi^\perp(z,\; \mu=1\; {\rm GeV})$ \cite{Straub:2015ica}.

%%%%%%%%%%%%%%%%%%%%%%%%%%%%%%%%%%%%%%%%%
\section{$B\rar K^*$ transition form factors}
% % % % % % % % % % % % % % % % % % % % %
  As we noted before, the form factors, computed via LCSR, are valid at low to intermediate $q^2$.  The extrapolation to high $q^2$ is performed via a two-parameter fit of the following form
%%%%%%%%%%%%%%%%%%%%%%%%%%%%%%%%%%%%%%%%%%
\begin{equation}
F(q^2)=\frac{F(0)}{1-a(q^2/m_B^2)+b(q^4/m_B^4)}
\end{equation}
to the LCSR predictions as well as form factor values obtained by the lattice QCD which are available at high $q^2$.  The results for the above fit are summarized in Table \ref{table:formfitslattice}.   \noindent $A_{12}$ is a combination of the two form factors $A_1$ and $A_2$  which appears in the expression for the differential decay rate and  is given as:
\be
A_{12}(q^2)=\frac{(m_B +m_{K^*})^2(m_{B}^2 -m_{K^*}^2-q^2)A_1(q^2)-\lambda_{K^*}(q^2)A_2(q^2)}{16m_Bm_{K^*}^2(m_B + m_{K^*})} \; ,\label{a12}
\ee
where $\lambda_{K^*}$ is a kinematical factor and is given as the following:
	\bea
	\lambda_{K^*}(q^2)&=m_B^4+m_{K^*}^4+q^4-2(m_B^2m_{K^*}^2+m_B^2q^2+m_{K^*}^2q^2)\; .\nonumber
	\eea
%%%%%%%%%%%%%%%%%%%%%%%%%%%%%%%%%%%%%%%%%%

\begin{table}[h]
	\begin{tabular}{|c|c|c|c|c|c|c|}
		\hline  & F(0) (LFHQCD)& F(0) (SR) & a (LFHQCD) & a (SR) & b (LFHQCD) & b (SR) \\ 
		\hline $V$ &$ 0.38^{+0.01}_{-0.03}$ & $0.43\pm 0.03$& $1.53^{+0.09}_{-0.05}$ & $1.67^{+0.11}_{-0.10}$& $0.62^{+0.14}_{-0.12}$ & $0.90^{+0.13}_{-0.11}$\\ 
		\hline $A_1$ & $0.29^{+0.01}_{-0.02}$ &$0.34\pm 0.02$ & $0.24^{+0.11}_{-0.06}$ &$0.36\pm 0.17$ & $-0.68^{+0.18}_{-0.16}$ & $-0.37\pm 0.17$\\ 
		\hline $A_{12}$ & $0.21\pm 0.01$ & $0.25\pm 0.01$& $0.33^{+0.08}_{-0.07}$ & $0.11^{+0.15}_{-0.14}$&$ -0.56^{+0.16}_{-0.15}$ & $-0.61\pm 0.12$\\ 
		\hline  
	\end{tabular}
	\caption{LFHQCD+ lattice prediction for the form factors. Lattice data is taken from \cite{Horgan:2013hoa}.  The error bars for the holographic form factors are due to the variation of the quark masses as explained in the text.}
	\label{table:formfitslattice}
\end{table}  

%%%%%%%%%%%%%%%%%%%%%%%%%%%%%%%%%%%%%%%%%%%

\begin{figure}[htbp]
	\begin{subfigure}{}
	\centering
	\includegraphics[width=0.3\textwidth]{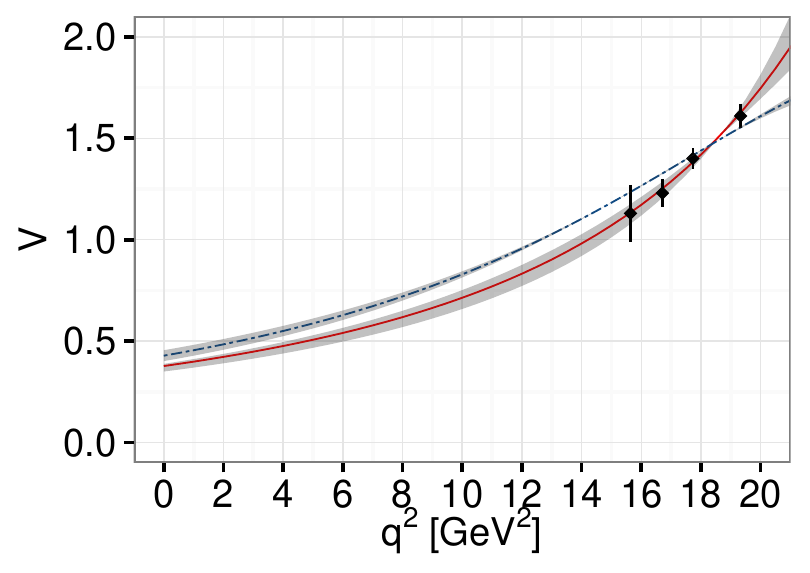}
	%\caption{1a}
	\label{fig:plot_V}
\end{subfigure}
\hspace{.1cm}
\begin{subfigure}{}
	\centering
	\includegraphics[width=0.3\textwidth]{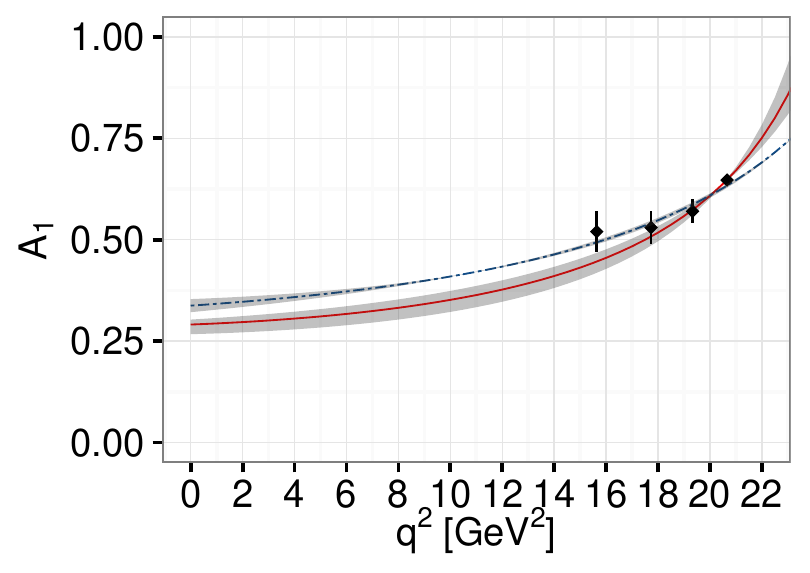}
	%\caption{1b}
	\label{fig:plot_A1}
\end{subfigure}
\begin{subfigure}{}
	\centering
	\includegraphics[width=0.3\textwidth]{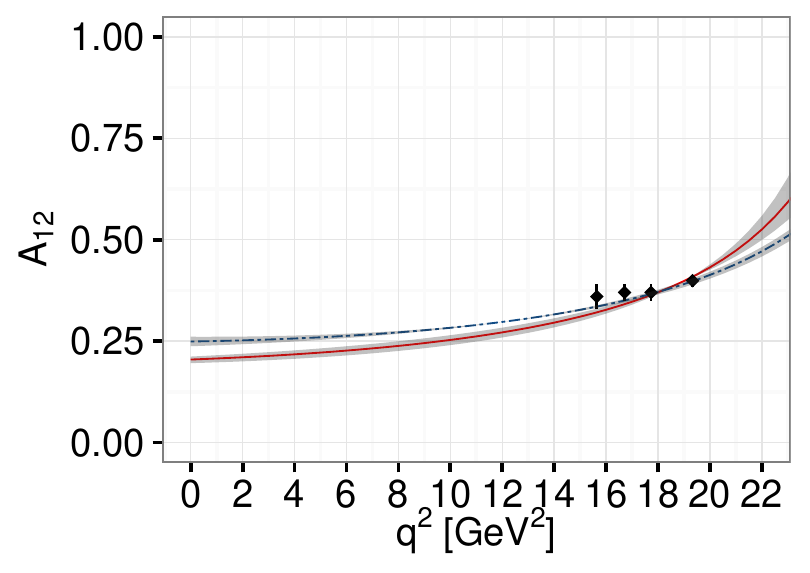}
	%\caption{1c}
	\label{fig:plot_A12}
	\end{subfigure}
	\caption{LFHQCD predictions for the form factors $V$, $A_1$ and $A_{12}$. The two-parameter fits with the available lattice data (red) are shown and compared with the predictions of QCD SM (dashed blue).  The shaded band represents the uncertainty in the predicted form factors due to uncertainty bands in DAs and variation in quark masses.}
	\label{formfactors}
\end{figure}

  Figures \ref{formfactors} shows the LFHQCD predictions including the lattice data points at high $q^2$ for the form factors $V$, $A_1$ and $A_{12}$.  The shaded bands in these figures represent the uncertainty due to the error band in the DAs.  Note that there is an additional uncertainly in the form factors inherent in the LCSR method (uncertainty in the Borel parameter, continuum threshold and other input parameters).  Since our goal in this paper is to discriminate between the LFHQCD and SR models and that the inherent LCSR uncertainties are the same in both models, we do not include them here.
  %%%%%%%%%%%%%%%%%%%%%%%%%%%
	Table \ref{inputs} shows the numerical values of the input parameters used in our predictions of the form factors and the decay rate.
	\begin{table}[h]
	\begin{tabular}{|c|c|c|c|}
		\hline $s^2_w$ & $0.23126(5)$ & $M_{B}$ & $8\;\mathrm{GeV}^2$   \\ 
		\hline $\alpha$ & $127.925(16)$ & $s_0$ & $36\;\mathrm{GeV}^2$   \\ 
		\hline	$|V_{tb}V_{ts}^{*}|$ & $0.0407(10)$ & $f_B$ & $0.18(1)\;\mathrm{GeV}$   \\ 
		\hline	$\tau_{B}$ & $1.519(5) ps$ & $m^{1S}_b$ & $4.60\;\mathrm{ GeV}$   \\  
		\hline 
		\end{tabular}
		\caption{Numerical values of the input parameters.}
		\label{inputs}
	\end{table}
%%%%%%%%%%%%%%%%%%%%%%%%%%%%%%%%%%%%%%%%%%%%%%%%%%%

%%%%%%%%%%%%%%%%%%%%%%%%%%%%%%%%

%%%%%%%%%%%%%%%%%%%%%%%%%%% 	
\section{Differential decay rate}
%%%%%%%%%%%%%%%%%%%%%%%%%%%
Once the form factors are known, the differential branching ratio for $B\to K^*\nu\bar\nu$ can be written as\cite{Buras:2014fpa}:
\bea
%\frac{dBR(\processK)}{dq^2}&=&\tau_{B^{+}}3\lvert N \rvert^2\frac{X^2}{s_w^4}\rho_{K}(q^2)\; ,\\
\frac{d\mathcal{BR}(\process)}{dq^2}&=&\tau_{B^0}3\lvert N \rvert^2\frac{X^2}{s_w^4}[\rho_{A_1}(q^2)+\rho_{A_{12}}(q^2)+\rho_{V}(q^2)] \; ,
\label{DBR}
\eea
where
	\be
	N=V_{tb}V_{ts}^{*}\frac{G_F\alpha }{16\pi^2}\sqrt{\frac{m_B}{3\pi}} \; ,\nonumber
	\ee
 
 and the functions $\rho_V$, $\rho_{A_1}$ and $\rho_{A_{12}}$ are defined in terms of the form factors:

	\bea
%	\rho_{K}(q^2)&=&\frac{\lambda_{K}^{3/2}(q^2)}{m_B^4}[f^{K}_+(q^2)]^2\; ,\label{rhok}
%	\\
		\rho_{V}(q^2)&=&\frac{2q^2\lambda_{K^*}^{3/2}(q^2)}{(m_B+m_{K^*})^2m_B^4}[V(q^2)]^2\; ,\label{rhov}
		\\
		\rho_{A_1}(q^2)&=&\frac{2q^2\lambda_{K^*}^{1/2}(q^2)(m_B+m_{K^*})^2}{m_B^4}[A_1(q^2)]^2\; ,\label{rhoa1}
		\\
		\rho_{A_{12}}(q^2)&=&\frac{64m_{K^*}^2\lambda_{K^*}^{1/2}(q^2)}{m_B^2}[A_{12}(q^2)]^2\; .\label{rhoa12}
	\eea

\begin{figure}[htbp]
	\centering
	\includegraphics[width=1\textwidth]{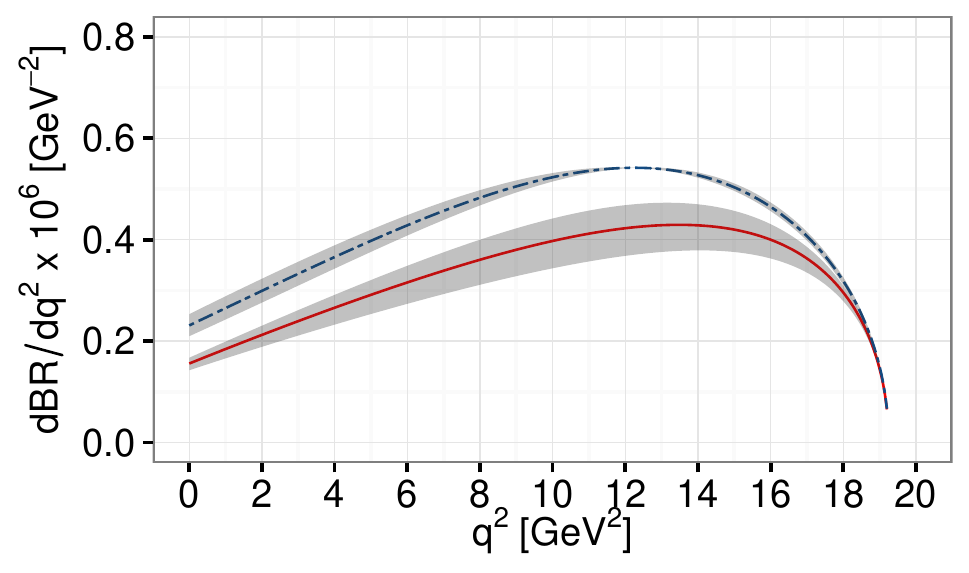}
	\caption{The LFHQCD (Solid line) and SR (Dashed line) predictions for the differential Branching Ratio for \process.  The shaded band represents the uncertainty coming from the form factors.}
	\label{fig:plot_BRKstar}
\end{figure}

	Figure \ref{fig:plot_BRKstar} compare the LFHQCD and SR predictions for the \process differential decay rate.  The resulting uncertainty due to form factors is shown as the shaded band.  We observe that, in general,  the LFHQCD prediction is lower than SR prediction for all values of the momentum transfer $q^2$.  The difference between the two predictions is maximal ( $\sim 25\%$) for intermediate values of $q^2$.  Most interestingly, the two predictions are quite distinct at low-to-intermediate $q^2$ where LCSR method is most reliable.  We expect that a future measurement of this decay channel at BELLE II may be able to discriminate between the two predictions.  For the total branching ratio, we predict $\mathcal{BR}(\process)_{\rm AdS/QCD}=(6.36^{+0.59}_{-0.74})\times 10^{-6}$, compared to sum rules result $\mathcal{BR}(\process)_{\rm SR}=(8.14^{+0.16}_{- 0.17})\times 10^{-6}$.

The $K^*$ longitudinal polarization fraction $F_L$ is another observable associated with \process decay.  Indeed, within the SM, the branching ratio for B decay to longitudinal $K^*$ and the neutrino-anti-neutrino pair is due to the second term in Eq. \ref{DBR} and therefore, for a given $q^2$, the fraction $F_L$ can be written as \cite{Altmannshofer:2009ma}:
		\be
	\frac{d	F_L(\process)_{SM}}{dq^2}=\frac{\rho_{A_{12}}(q^2)}{\rho_{A_1}(q^2)+\rho_{A_{12}}(q^2)+\rho_{V}(q^2)}\; .
		\ee

\begin{figure}[htbp]
	\centering
	\includegraphics[width=1\textwidth]{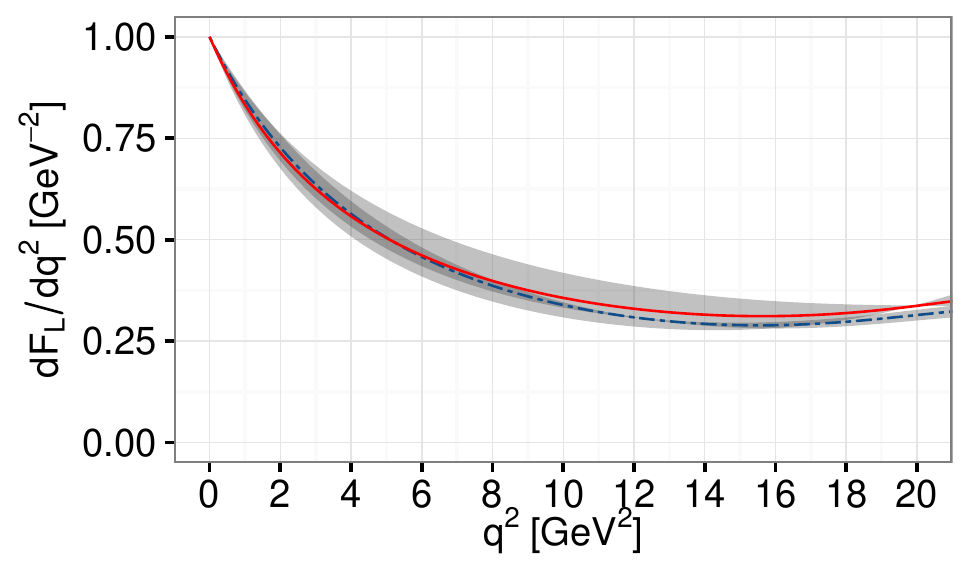}
	\caption{The LFHQCD (Solid line) and SR (Dashed line) predictions for the polarization fraction distribution for \process.  The shaded band represents the uncertainty coming from the form factors.}
	\label{fig:plot_FL}
\end{figure}	
Figure \ref{fig:plot_FL} shows our predictions for $F_L$ versus $q^2$.  We observe that within error bands, the two model predictions are not distinguishable.  This confirms that the $K^*$ longitudinal polarization fraction $F_L$ has little sensitivity to the non-perturbative form factors and is thus an excellent observable to probe New Physics signals.  Integrating over the whole kinematic region $0\le q^2\le (m_B-m_{K^*})^2$, we predict
\be
 F_L(\process)_{\rm SM}^{\rm LFHQCD}=0.40^{+0.02}_{-0.01} \;\; ,
\label{ourSMpredictions}
\ee
as compared to $F_L(\process)_{\rm SM}^{\rm SR}=0.41\pm 0.01$.

In Table \ref{table:databin}, we present bin-by bin predictions of LFHQCD and SR for the branching ratio and longitudinal polarization asymmetry.  The predictions in a $q^2$ bin $[a,b]$ are computed using
\be
\langle \mathcal{BR}^{}_{K^{*}}\rangle \int_a^b \frac{d\mathcal{BR}(\process)}{dq^2}dq^2 \; ,
\ee
and
\be
\langle F_L^{}\rangle  \frac{\int_a^b \rho_{A_{12}}(q^2)dq^2}{\int_a^b (\rho_{A_1}(q^2)+\rho_{A_{12}}(q^2)+\rho_{V}(q^2)) dq^2 }\; .
\ee
\begin{table}[h]
	\begin{tabular}{|ccccc|}
		\hline $q^2$ & $10^6 \langle \mathcal{BR}^{LFHQCD}\rangle$ & $10^6 \langle \mathcal{BR}^{SR}\rangle$ & $\langle F_L^{LFHQCD}\rangle$& $\langle F_L^{SR}\rangle$  \\ 
		\hline 
		$0-4$ & $0.85^{+0.07}_{-0.10}$ &$1.20{\pm}0.09$& $0.71^{+0.002}_{-0.01}$&$0.73\pm0.01$ \\ 
		$4-8$ & $1.26^{+0.13}_{-0.17}$ &$1.71{\pm}0.08$& $0.46^{+0.02}_{-0.01}$&$0.46\pm 0.01$   \\ 
		$8-12$ & $1.58^{+0.18}_{-0.21}$ &$2.08{\pm}0.03$& $0.35_{-0.01}^{+0.02}$&$0.35\pm 0.01$  \\ 
		$12-16$ & $1.69^{+0.16}_{-0.19}$ &$2.08{\pm}0.02$& $0.32\pm {0.01}$&$0.30\pm 0.01$   \\ 
		$16-max$ & $0.97\pm{0.06}$ &$1.08^{+0.02}_{-0.03}$& $0.32\pm{0.01}$&$0.30\pm 0.01$ \\ 
		\hline
		$0-max$ & $6.36^{+0.59}_{-0.74}$ &$8.14^{+0.16}_{-0.17}$& $0.40_{-0.01}^{+0.02}$&$0.41\pm 0.01$   \\ 
		\hline
	\end{tabular}
\caption{Bin-by-bin LFHQCD and SR predictions for the branching ratio and longitudinal polarization asymmetry.}
\label{table:databin}
\end{table}

%%%%%%%%%%%%%%%%%%%%%	
\section{Conclusion}%
%%%%%%%%%%%%%%%%%%%%%

Experimental observation of \process can provide an excellent probe for the theoretical $B\to K^*$ transition form factors.  The differential branching ratio for this decay shows the largest sensitivity to the form factors for low-to-intermediate values of the momentum transfer.  The $K^*$ longitudinal polarization fraction, on the other hand, is not sensitive to the form factors, which makes it an interesting observable for NP search. 

%%%%%%%%%%%%%%%%%%%%%%%%%%
\section{Acknowledgement}
%%%%%%%%%%%%%%%%%%%%%%%%%
M. A. and R. S. are supported by individual Discovery Grants from the Natural Sciences and Engineering Research Council of Canada (NSERC): No. SAPIN-2017-00033 and No. SAPIN-2017-00031, respectively.  We thank Michael Thibodeau for his input in coding.

\bibliographystyle{apsrev}
\bibliography{BKstarnu}

\begin{thebibliography}{29}
\expandafter\ifx\csname natexlab\endcsname\relax\def\natexlab#1{#1}\fi
\expandafter\ifx\csname bibnamefont\endcsname\relax
  \def\bibnamefont#1{#1}\fi
\expandafter\ifx\csname bibfnamefont\endcsname\relax
  \def\bibfnamefont#1{#1}\fi
\expandafter\ifx\csname citenamefont\endcsname\relax
  \def\citenamefont#1{#1}\fi
\expandafter\ifx\csname url\endcsname\relax
  \def\url#1{\texttt{#1}}\fi
\expandafter\ifx\csname urlprefix\endcsname\relax\def\urlprefix{URL }\fi
\providecommand{\bibinfo}[2]{#2}
\providecommand{\eprint}[2][]{\url{#2}}

\bibitem[{\citenamefont{Lutz et~al.}(2013)}]{Lutz:2013ftz}
\bibinfo{author}{\bibfnamefont{O.}~\bibnamefont{Lutz}} \bibnamefont{et~al.}
  (\bibinfo{collaboration}{Belle}), \bibinfo{journal}{Phys. Rev.}
  \textbf{\bibinfo{volume}{D87}}, \bibinfo{pages}{111103}
  (\bibinfo{year}{2013}), \eprint{1303.3719}.

\bibitem[{\citenamefont{Aushev et~al.}(2010)}]{Aushev:2010bq}
\bibinfo{author}{\bibfnamefont{T.}~\bibnamefont{Aushev}} \bibnamefont{et~al.}
  (\bibinfo{year}{2010}), \eprint{1002.5012}.

\bibitem[{\citenamefont{Ahmady et~al.}(2014{\natexlab{a}})\citenamefont{Ahmady,
  Campbell, Lord, and Sandapen}}]{Ahmady:2014sva}
\bibinfo{author}{\bibfnamefont{M.}~\bibnamefont{Ahmady}},
  \bibinfo{author}{\bibfnamefont{R.}~\bibnamefont{Campbell}},
  \bibinfo{author}{\bibfnamefont{S.}~\bibnamefont{Lord}}, \bibnamefont{and}
  \bibinfo{author}{\bibfnamefont{R.}~\bibnamefont{Sandapen}},
  \bibinfo{journal}{Phys. Rev.} \textbf{\bibinfo{volume}{D89}},
  \bibinfo{pages}{074021} (\bibinfo{year}{2014}{\natexlab{a}}),
  \eprint{1401.6707}.

\bibitem[{\citenamefont{Bharucha et~al.}(2016)\citenamefont{Bharucha, Straub,
  and Zwicky}}]{Straub:2015ica}
\bibinfo{author}{\bibfnamefont{A.}~\bibnamefont{Bharucha}},
  \bibinfo{author}{\bibfnamefont{D.~M.} \bibnamefont{Straub}},
  \bibnamefont{and} \bibinfo{author}{\bibfnamefont{R.}~\bibnamefont{Zwicky}},
  \bibinfo{journal}{JHEP} \textbf{\bibinfo{volume}{08}}, \bibinfo{pages}{098}
  (\bibinfo{year}{2016}), \eprint{1503.05534}.

\bibitem[{\citenamefont{Adloff et~al.}(2000)}]{Adloff:1999kg}
\bibinfo{author}{\bibfnamefont{C.}~\bibnamefont{Adloff}} \bibnamefont{et~al.}
  (\bibinfo{collaboration}{H1}), \bibinfo{journal}{Eur. Phys. J.}
  \textbf{\bibinfo{volume}{C13}}, \bibinfo{pages}{371} (\bibinfo{year}{2000}),
  \eprint{hep-ex/9902019}.

\bibitem[{\citenamefont{Aid et~al.}(1996)}]{Aid:1996bs}
\bibinfo{author}{\bibfnamefont{S.}~\bibnamefont{Aid}} \bibnamefont{et~al.}
  (\bibinfo{collaboration}{H1}), \bibinfo{journal}{Nucl. Phys.}
  \textbf{\bibinfo{volume}{B463}}, \bibinfo{pages}{3} (\bibinfo{year}{1996}),
  \eprint{hep-ex/9601004}.

\bibitem[{\citenamefont{Breitweg et~al.}(1998)}]{Breitweg:1997ed}
\bibinfo{author}{\bibfnamefont{J.}~\bibnamefont{Breitweg}} \bibnamefont{et~al.}
  (\bibinfo{collaboration}{ZEUS}), \bibinfo{journal}{Eur. Phys. J.}
  \textbf{\bibinfo{volume}{C2}}, \bibinfo{pages}{247} (\bibinfo{year}{1998}),
  \eprint{hep-ex/9712020}.

\bibitem[{\citenamefont{Chekanov et~al.}(2005)}]{Chekanov:2005cqa}
\bibinfo{author}{\bibfnamefont{S.}~\bibnamefont{Chekanov}} \bibnamefont{et~al.}
  (\bibinfo{collaboration}{ZEUS}), \bibinfo{journal}{Nucl. Phys.}
  \textbf{\bibinfo{volume}{B718}}, \bibinfo{pages}{3} (\bibinfo{year}{2005}),
  \eprint{hep-ex/0504010}.

\bibitem[{\citenamefont{{The H1 Collaboration}}(2010)}]{Collaboration:2009xp}
\bibinfo{author}{\bibnamefont{{The H1 Collaboration}}}, \bibinfo{journal}{JHEP}
  \textbf{\bibinfo{volume}{05}}, \bibinfo{pages}{032} (\bibinfo{year}{2010}),
  \eprint{0910.5831}.

\bibitem[{\citenamefont{Chekanov et~al.}(2007)}]{Chekanov:2007zr}
\bibinfo{author}{\bibfnamefont{S.}~\bibnamefont{Chekanov}} \bibnamefont{et~al.}
  (\bibinfo{collaboration}{ZEUS}), \bibinfo{journal}{PMC Phys.}
  \textbf{\bibinfo{volume}{A1}}, \bibinfo{pages}{6} (\bibinfo{year}{2007}),
  \eprint{0708.1478}.

\bibitem[{\citenamefont{Forshaw and Sandapen}(2012)}]{Forshaw:2012im}
\bibinfo{author}{\bibfnamefont{J.~R.} \bibnamefont{Forshaw}} \bibnamefont{and}
  \bibinfo{author}{\bibfnamefont{R.}~\bibnamefont{Sandapen}},
  \bibinfo{journal}{Phys.Rev.Lett.} \textbf{\bibinfo{volume}{109}},
  \bibinfo{pages}{081601} (\bibinfo{year}{2012}), \eprint{1203.6088}.

\bibitem[{\citenamefont{Ahmady et~al.}(2016)\citenamefont{Ahmady, Sandapen, and
  Sharma}}]{Ahmady:2016ujw}
\bibinfo{author}{\bibfnamefont{M.}~\bibnamefont{Ahmady}},
  \bibinfo{author}{\bibfnamefont{R.}~\bibnamefont{Sandapen}}, \bibnamefont{and}
  \bibinfo{author}{\bibfnamefont{N.}~\bibnamefont{Sharma}},
  \bibinfo{journal}{Phys. Rev.} \textbf{\bibinfo{volume}{D94}},
  \bibinfo{pages}{074018} (\bibinfo{year}{2016}), \eprint{1605.07665}.

\bibitem[{\citenamefont{Ahmady et~al.}(2013)\citenamefont{Ahmady, Campbell,
  Lord, and Sandapen}}]{Ahmady:2013cga}
\bibinfo{author}{\bibfnamefont{M.}~\bibnamefont{Ahmady}},
  \bibinfo{author}{\bibfnamefont{R.}~\bibnamefont{Campbell}},
  \bibinfo{author}{\bibfnamefont{S.}~\bibnamefont{Lord}}, \bibnamefont{and}
  \bibinfo{author}{\bibfnamefont{R.}~\bibnamefont{Sandapen}},
  \bibinfo{journal}{Phys. Rev.} \textbf{\bibinfo{volume}{D88}},
  \bibinfo{pages}{074031} (\bibinfo{year}{2013}), \eprint{1308.3694}.

\bibitem[{\citenamefont{Ahmady et~al.}(2014{\natexlab{b}})\citenamefont{Ahmady,
  Lord, and Sandapen}}]{Ahmady:2014cpa}
\bibinfo{author}{\bibfnamefont{M.}~\bibnamefont{Ahmady}},
  \bibinfo{author}{\bibfnamefont{S.}~\bibnamefont{Lord}}, \bibnamefont{and}
  \bibinfo{author}{\bibfnamefont{R.}~\bibnamefont{Sandapen}},
  \bibinfo{journal}{Phys.Rev.} \textbf{\bibinfo{volume}{D90}},
  \bibinfo{pages}{074010} (\bibinfo{year}{2014}{\natexlab{b}}),
  \eprint{1407.6700}.

\bibitem[{\citenamefont{Ahmady et~al.}(2015)\citenamefont{Ahmady, Hatfield,
  Lord, and Sandapen}}]{Ahmady:2015fha}
\bibinfo{author}{\bibfnamefont{M.}~\bibnamefont{Ahmady}},
  \bibinfo{author}{\bibfnamefont{D.}~\bibnamefont{Hatfield}},
  \bibinfo{author}{\bibfnamefont{S.}~\bibnamefont{Lord}}, \bibnamefont{and}
  \bibinfo{author}{\bibfnamefont{R.}~\bibnamefont{Sandapen}},
  \bibinfo{journal}{Phys. Rev.} \textbf{\bibinfo{volume}{D92}},
  \bibinfo{pages}{114028} (\bibinfo{year}{2015}), \eprint{1508.02327}.

\bibitem[{\citenamefont{Buras et~al.}(2015)\citenamefont{Buras, Girrbach-Noe,
  Niehoff, and Straub}}]{Buras:2014fpa}
\bibinfo{author}{\bibfnamefont{A.~J.} \bibnamefont{Buras}},
  \bibinfo{author}{\bibfnamefont{J.}~\bibnamefont{Girrbach-Noe}},
  \bibinfo{author}{\bibfnamefont{C.}~\bibnamefont{Niehoff}}, \bibnamefont{and}
  \bibinfo{author}{\bibfnamefont{D.~M.} \bibnamefont{Straub}},
  \bibinfo{journal}{JHEP} \textbf{\bibinfo{volume}{02}}, \bibinfo{pages}{184}
  (\bibinfo{year}{2015}), \eprint{1409.4557}.

\bibitem[{\citenamefont{Olive et~al.}(2014)}]{Agashe:2014kda}
\bibinfo{author}{\bibfnamefont{K.~A.} \bibnamefont{Olive}} \bibnamefont{et~al.}
  (\bibinfo{collaboration}{Particle Data Group}), \bibinfo{journal}{Chin.
  Phys.} \textbf{\bibinfo{volume}{C38}}, \bibinfo{pages}{090001}
  (\bibinfo{year}{2014}).

\bibitem[{\citenamefont{Misiak and Urban}(1999)}]{Misiak:1999yg}
\bibinfo{author}{\bibfnamefont{M.}~\bibnamefont{Misiak}} \bibnamefont{and}
  \bibinfo{author}{\bibfnamefont{J.}~\bibnamefont{Urban}},
  \bibinfo{journal}{Phys. Lett.} \textbf{\bibinfo{volume}{B451}},
  \bibinfo{pages}{161} (\bibinfo{year}{1999}), \eprint{hep-ph/9901278}.

\bibitem[{\citenamefont{Buchalla and Buras}(1999)}]{Buchalla:1998ba}
\bibinfo{author}{\bibfnamefont{G.}~\bibnamefont{Buchalla}} \bibnamefont{and}
  \bibinfo{author}{\bibfnamefont{A.~J.} \bibnamefont{Buras}},
  \bibinfo{journal}{Nucl. Phys.} \textbf{\bibinfo{volume}{B548}},
  \bibinfo{pages}{309} (\bibinfo{year}{1999}), \eprint{hep-ph/9901288}.

\bibitem[{\citenamefont{Brod et~al.}(2011)\citenamefont{Brod, Gorbahn, and
  Stamou}}]{Brod:2010hi}
\bibinfo{author}{\bibfnamefont{J.}~\bibnamefont{Brod}},
  \bibinfo{author}{\bibfnamefont{M.}~\bibnamefont{Gorbahn}}, \bibnamefont{and}
  \bibinfo{author}{\bibfnamefont{E.}~\bibnamefont{Stamou}},
  \bibinfo{journal}{Phys. Rev.} \textbf{\bibinfo{volume}{D83}},
  \bibinfo{pages}{034030} (\bibinfo{year}{2011}), \eprint{1009.0947}.

\bibitem[{\citenamefont{Ahmady and Sandapen}(2013)}]{Ahmady:2013cva}
\bibinfo{author}{\bibfnamefont{M.}~\bibnamefont{Ahmady}} \bibnamefont{and}
  \bibinfo{author}{\bibfnamefont{R.}~\bibnamefont{Sandapen}},
  \bibinfo{journal}{Phys.Rev.D} \textbf{\bibinfo{volume}{88}},
  \bibinfo{pages}{014042} (\bibinfo{year}{2013}), \eprint{1305.1479}.

\bibitem[{\citenamefont{Brodsky et~al.}(2014)\citenamefont{Brodsky,
  de~Teramond, Dosch, and Erlich}}]{Brodsky:2014yha}
\bibinfo{author}{\bibfnamefont{S.~J.} \bibnamefont{Brodsky}},
  \bibinfo{author}{\bibfnamefont{G.~F.} \bibnamefont{de~Teramond}},
  \bibinfo{author}{\bibfnamefont{H.~G.} \bibnamefont{Dosch}}, \bibnamefont{and}
  \bibinfo{author}{\bibfnamefont{J.}~\bibnamefont{Erlich}}
  (\bibinfo{year}{2014}), \eprint{1407.8131}.

\bibitem[{\citenamefont{Beringer et~al.}(2012)}]{Beringer:1900zz}
\bibinfo{author}{\bibfnamefont{J.}~\bibnamefont{Beringer}} \bibnamefont{et~al.}
  (\bibinfo{collaboration}{Particle Data Group}), \bibinfo{journal}{Phys.Rev.}
  \textbf{\bibinfo{volume}{D86}}, \bibinfo{pages}{010001}
  (\bibinfo{year}{2012}).

\bibitem[{\citenamefont{Becirevic et~al.}(2003)\citenamefont{Becirevic, Lubicz,
  Mescia, and Tarantino}}]{Becirevic:2003pn}
\bibinfo{author}{\bibfnamefont{D.}~\bibnamefont{Becirevic}},
  \bibinfo{author}{\bibfnamefont{V.}~\bibnamefont{Lubicz}},
  \bibinfo{author}{\bibfnamefont{F.}~\bibnamefont{Mescia}}, \bibnamefont{and}
  \bibinfo{author}{\bibfnamefont{C.}~\bibnamefont{Tarantino}},
  \bibinfo{journal}{JHEP} \textbf{\bibinfo{volume}{0305}}, \bibinfo{pages}{007}
  (\bibinfo{year}{2003}), \eprint{hep-lat/0301020}.

\bibitem[{\citenamefont{Braun et~al.}(2003)\citenamefont{Braun, Burch,
  Gattringer, Gockeler, Lacagnina et~al.}}]{Braun:2003jg}
\bibinfo{author}{\bibfnamefont{V.}~\bibnamefont{Braun}},
  \bibinfo{author}{\bibfnamefont{T.}~\bibnamefont{Burch}},
  \bibinfo{author}{\bibfnamefont{C.}~\bibnamefont{Gattringer}},
  \bibinfo{author}{\bibfnamefont{M.}~\bibnamefont{Gockeler}},
  \bibinfo{author}{\bibfnamefont{G.}~\bibnamefont{Lacagnina}},
  \bibnamefont{et~al.}, \bibinfo{journal}{Phys.Rev.}
  \textbf{\bibinfo{volume}{D68}}, \bibinfo{pages}{054501}
  (\bibinfo{year}{2003}), \eprint{hep-lat/0306006}.

\bibitem[{\citenamefont{Ball et~al.}(2007)\citenamefont{Ball, Braun, and
  Lenz}}]{Ball:2007zt}
\bibinfo{author}{\bibfnamefont{P.}~\bibnamefont{Ball}},
  \bibinfo{author}{\bibfnamefont{V.~M.} \bibnamefont{Braun}}, \bibnamefont{and}
  \bibinfo{author}{\bibfnamefont{A.}~\bibnamefont{Lenz}},
  \bibinfo{journal}{JHEP} \textbf{\bibinfo{volume}{08}}, \bibinfo{pages}{090}
  (\bibinfo{year}{2007}), \eprint{0707.1201}.

\bibitem[{\citenamefont{Choi and Ji}(2007)}]{Choi:2007yu}
\bibinfo{author}{\bibfnamefont{H.-M.} \bibnamefont{Choi}} \bibnamefont{and}
  \bibinfo{author}{\bibfnamefont{C.-R.} \bibnamefont{Ji}},
  \bibinfo{journal}{Phys. Rev.} \textbf{\bibinfo{volume}{D75}},
  \bibinfo{pages}{034019} (\bibinfo{year}{2007}), \eprint{hep-ph/0701177}.

\bibitem[{\citenamefont{Horgan et~al.}(2014)\citenamefont{Horgan, Liu, Meinel,
  and Wingate}}]{Horgan:2013hoa}
\bibinfo{author}{\bibfnamefont{R.~R.} \bibnamefont{Horgan}},
  \bibinfo{author}{\bibfnamefont{Z.}~\bibnamefont{Liu}},
  \bibinfo{author}{\bibfnamefont{S.}~\bibnamefont{Meinel}}, \bibnamefont{and}
  \bibinfo{author}{\bibfnamefont{M.}~\bibnamefont{Wingate}},
  \bibinfo{journal}{Phys. Rev.} \textbf{\bibinfo{volume}{D89}},
  \bibinfo{pages}{094501} (\bibinfo{year}{2014}), \eprint{1310.3722}.

\bibitem[{\citenamefont{Altmannshofer et~al.}(2009)\citenamefont{Altmannshofer,
  Buras, Straub, and Wick}}]{Altmannshofer:2009ma}
\bibinfo{author}{\bibfnamefont{W.}~\bibnamefont{Altmannshofer}},
  \bibinfo{author}{\bibfnamefont{A.~J.} \bibnamefont{Buras}},
  \bibinfo{author}{\bibfnamefont{D.~M.} \bibnamefont{Straub}},
  \bibnamefont{and} \bibinfo{author}{\bibfnamefont{M.}~\bibnamefont{Wick}},
  \bibinfo{journal}{JHEP} \textbf{\bibinfo{volume}{04}}, \bibinfo{pages}{022}
  (\bibinfo{year}{2009}), \eprint{0902.0160}.

\end{thebibliography}

\end{document}